\def\realsNonnegative{\mathbb{R}_{0}^{+}}
\def\realsNonnegative{\mathbb{R}_{0}^{+}}
\def\complexNumbers{\mathbb{C}}
\def\realNumbers{\mathbb{R}}
\def\integers{\mathbb{Z}}
\def\integersPositive{\mathbb{Z}^{+}}
\def\integersNonnegative{\mathbb{Z}_{0}^{+}}
\def\generalSet{\mathbb{S}}

\def\constante{{\rm e}}
\def\constantj{{\rm j}}

\def\lagForCorrelation{k}

\def\indexIteration{n}

\def\indexIterationANF{l}

\def\indexFirstOrderMonomial{j}
\def\indexMonomial{k}
\def\orderMonomial[#1]{k_{#1}}
\def\coeffientsANF[#1]{c_{#1}}
\def\polyVariable{z}

\def\numberOfIterations{m}
\def\numberOfPointsForPSK{H}
\def\modulationSymbolF[#1]{m_{#1}}
\def\cardinalitySetOfOperators[#1]{{H}_{#1}}

\def\varMonomial{x}
\def\monomial[#1]{x_{#1}}

\def\lengthGaGb{N}

\def\scaleAexp[#1]{a_{#1}}
\def\scaleBexp[#1]{b_{#1}}
\def\scaleEexp[#1]{e_{#1}}
\def\angleexp[#1]{c_{#1}}
\def\angleexpAll[#1]{k_{#1}}
\def\angleScaleAexp[#1]{\dot{c}_{#1}}
\def\angleScaleBexp[#1]{\ddot{c}_{#1}}
\def\arbitraryScaleE{e_0}

\def\arbitraryPhaseK{k_0}

\def\separationGolay[#1]{d_{#1}}

\def\lossFunction{J}
\def\setMiniBatch{\mathbb{B}}
\def\indexSampleInBatch{i}
\def\sizeofBatch{B}
\def\scaleEexpBatch[#1][#2]{e_{#1}^{(#2)}}
\def\angleexpAllBatch[#1][#2]{k_{#1}^{(#2)}}

\def\deviationPhase{\beta}
\def\deviationAmplitude{\alpha}
\def\scaleEexpPre[#1]{\dot{e}_{#1}}
\def\angleexpAllPre[#1]{\dot{k}_{#1}}

\def\eleGa[#1]{{a}_{#1}}
\def\eleGb[#1]{{b}_{#1}}

\def\apac[#1][#2]{\rho_{#1}(#2)}
\def\apacPositive[#1][#2]{\rho^{+}_{#1}(#2)}
\def\binaryAsignment[#1][#2]{b_{#1}^{(#2)}}

\def\eleSeqf[#1]{{f}_{#1}}
\def\eleSeqg[#1]{{g}_{#1}}
\def\eleSeqcf[#1]{{c}_{f,#1}}
\def\eleSeqcg[#1]{{c}_{g,#1}}

\def\scaleA[#1]{\alpha_{#1}}
\def\scaleB[#1]{\beta_{#1}}
\def\angleGolay[#1]{\omega_{#1}}
\def\angleScaleA[#1][#2]{\dot{\omega}_{#1}^{#2}}
\def\angleScaleB[#1][#2]{\ddot{\omega}_{#1}^{#2}}
\def\permutationShift[#1]{{\psi_{#1}}}
\def\permutationMono[#1]{{p_{#1}}}

\def\seqPermutationCompShift{\bm{p}}

\def\symbolDuration{T_{\rm s}}
\def\timeVar{t}

\def\funczArbitrary[#1]{z(#1)}
\def\funcaArbitrary[#1]{a(#1)}
\def\funcbArbitrary[#1]{b(#1)}
\def\coefficientArbitrary[#1]{k_{#1}}

\def\distanceToPoint[#1]{d_{#1}}
\def\ratioBetweenDistanceAndInner[#1]{l_{#1}}
\def\angleBetweenPointAndXaxis[#1]{\theta_{#1}}
\def\angleBetweenPointAndXYDiagonal[#1]{\psi_{#1}}
\def\angleBetweenPointAndSymPoint[#1]{\xi_{#1}}

\def\numberOfInfoBits{S}

\def\vecArrangement[#1]{\textbf{b}_{#1}}

\def\seqGa{\textit{\textbf{a}}}
\def\seqGb{\textit{\textbf{b}}}
\def\seqGaIt[#1]{\textit{\textbf{a}}^{(#1)}}
\def\seqGbIt[#1]{\textit{\textbf{b}}^{(#1)}}
\def\seqGc{\textit{\textbf{c}}}

\def\seqGf[#1]{\textit{\textbf{f}}_{#1}}
\def\seqGg[#1]{\textit{\textbf{g}}_{#1}}
\def\seqGfdot[#1]{\bar{\textit{\textbf{f}}}_{#1}}
\def\seqGgdot[#1]{\bar{\textit{\textbf{g}}}_{#1}}

\def\seqf{\textit{\textbf{f}}}

\def\seqSub[#1]{\textit{\textbf{h}}_{#1}}

\def\seqFirstOrderMonomial[#1]{\textit{\textbf{m}}_{#1}}
\def\seqx{\textit{\textbf{x}}}

\def\seqToBeModulated[#1]{\textit{\textbf{s}}_{#1}}

\def\flipConjugate[#1]{{{\tilde{#1}}}}
\def\expectationOperator[#1]{{\mathbb{E}}[#1]}
\def\operator[#1][#2]{\mathcal{O}_{#1}^{(#2)}}
\def\operatordot[#1][#2]{\bar{\mathcal{O}}_{#1}^{(#2)}}

\def\compositeOperatorF[#1][#2]{{F}_{#1}{(#2)}}
\def\compositeOperatorG[#1][#2]{{G}_{#1}{(#2)}}
\def\compositeOperatorFdot[#1][#2]{\bar{F}_{#1}{(#2)}}
\def\compositeOperatorGdot[#1][#2]{\bar{G}_{#1}{(#2)}}

\def\setOfOperators[#1]{{\mathfrak{J}}_{#1}}

\def\operatorBinary[#1][#2]{O_{#1}^{(#2)}}
\def\operatorSign[#1][#2]{{\rm S}_{#1}^{(#2)}}
\def\operatorScaleA[#1][#2]{{\rm{A}}_{#1}^{(#2)}}
\def\operatorScaleB[#1][#2]{{\rm{B}}_{#1}^{(#2)}}
\def\operatorAngle[#1][#2]{\Omega_{#1}^{(#2)}}
\def\operatorSeparation[#1][#2]{\Delta_{#1}^{(#2)}}
\def\operatorOrderA[#1][#2]{\dot{\rm O}_{#1}^{(#2)}}
\def\operatorOrderB[#1][#2]{\ddot{\rm O}_{#1}^{(#2)}}
\def\operatorAngleScaleA[#1][#2]{\dot{\Omega}_{#1}^{(#2)}}
\def\operatorAngleScaleB[#1][#2]{\ddot{\Omega}_{#1}^{(#2)}}
\def\operatorAngleConjScaleA[#1][#2]{\dot{\Upsilon}_{#1}^{(#2)}}
\def\operatorAngleConjScaleB[#1][#2]{\ddot{\Upsilon}_{#1}^{(#2)}}

\def\functionf[#1]{p^{(#1)}}
\def\functiong[#1]{q^{(#1)}}

\def\functionfdot[#1]{\bar{p}_{\indexIterationANF}^{(#1)}}
\def\functiongdot[#1]{\bar{q}_{\indexIterationANF}^{(#1)}}

\def\funcfForCommonShift{f_{\rm s}}
\def\funcfForCommonOrder{p_{\rm o}}

\def\funcfForFinalPhase{f_{\rm i}}
\def\funcfForFinalAmplitude{f_{\rm r}}
\def\funcgForFinalPhase{g_{\rm i}}

\def\funcfForANF{f}
\def\funcGfForANF[#1]{f_{#1}}
\def\funcGgForANF[#1]{g_{#1}}
\def\polySeq[#1][#2]{p_{#1}(#2)}

\newcommand\mydots{\hbox to 1em{.\hss.\hss.}}

\documentclass[journal]{IEEEtran}

\usepackage{multicol}
\usepackage{lipsum}
\usepackage{mathtools}
\usepackage{amsthm}
\usepackage{acronym}
\usepackage{stfloats}
\usepackage{amsfonts}
\usepackage{acronym}
\usepackage{cite}
\usepackage{multirow}
\usepackage{bm}
\usepackage{amsmath,amssymb}
\usepackage{graphicx}
\usepackage[table,xcdraw]{xcolor}
\usepackage[english]{babel}
\usepackage[caption=false,font=footnotesize]{subfig}
\usepackage[geometry]{ifsym}
\usepackage{array}
\usepackage[utf8]{inputenc}
\usepackage[T1]{fontenc}

\usepackage{tikz}
\usepackage{lipsum}
\usetikzlibrary{calc,shadings,patterns}
\makeatletter
\tikzset{%
  remember picture with id/.style={%
    remember picture,
    overlay,
    save picture id=#1,
  },
  save picture id/.code={%
    \edef\pgf@temp{#1}%
    \immediate\write\pgfutil@auxout{%
      \noexpand\savepointas{\pgf@temp}{\pgfpictureid}}%
  },
  if picture id/.code args={#1#2#3}{%
    \@ifundefined{save@pt@#1}{%
      \pgfkeysalso{#3}%
    }{
      \pgfkeysalso{#2}%
    }
  }
}

\def\savepointas#1#2{%
  \expandafter\gdef\csname save@pt@#1\endcsname{#2}%
}

\def\tmk@labeldef#1,#2\@nil{%
  \def\tmk@label{#1}%
  \def\tmk@def{#2}%
}

\tikzdeclarecoordinatesystem{pic}{%
  \pgfutil@in@,{#1}%
  \ifpgfutil@in@%
    \tmk@labeldef#1\@nil
  \else
    \tmk@labeldef#1,(0pt,0pt)\@nil
  \fi
  \@ifundefined{save@pt@\tmk@label}{%
    \tikz@scan@one@point\pgfutil@firstofone\tmk@def
  }{%
  \pgfsys@getposition{\csname save@pt@\tmk@label\endcsname}\save@orig@pic%
  \pgfsys@getposition{\pgfpictureid}\save@this@pic%
  \pgf@process{\pgfpointorigin\save@this@pic}%
  \pgf@xa=\pgf@x
  \pgf@ya=\pgf@y
  \pgf@process{\pgfpointorigin\save@orig@pic}%
  \advance\pgf@x by -\pgf@xa
  \advance\pgf@y by -\pgf@ya
  }%
}

\makeatother

\newcounter{hatchNumber}
\usepackage{cuted}
\makeatletter
\newif\ifAC@uppercase@first%
\def\Aclp#1{\AC@uppercase@firsttrue\aclp{#1}\AC@uppercase@firstfalse}%
\def\AC@aclp#1{%
	\ifcsname fn@#1@PL\endcsname%
	\ifAC@uppercase@first%
	\expandafter\expandafter\expandafter\MakeUppercase\csname fn@#1@PL\endcsname%
	\else%
	\csname fn@#1@PL\endcsname%
	\fi%
	\else%
	\AC@acl{#1}s%
	\fi%
}%
\def\Acp#1{\AC@uppercase@firsttrue\acp{#1}\AC@uppercase@firstfalse}%
\def\AC@acp#1{%
	\ifcsname fn@#1@PL\endcsname%
	\ifAC@uppercase@first%
	\expandafter\expandafter\expandafter\MakeUppercase\csname fn@#1@PL\endcsname%
	\else%
	\csname fn@#1@PL\endcsname%
	\fi%
	\else%
	\AC@ac{#1}s%
	\fi%
}%
\def\Acfp#1{\AC@uppercase@firsttrue\acfp{#1}\AC@uppercase@firstfalse}%
\def\AC@acfp#1{%
	\ifcsname fn@#1@PL\endcsname%
	\ifAC@uppercase@first%
	\expandafter\expandafter\expandafter\MakeUppercase\csname fn@#1@PL\endcsname%
	\else%
	\csname fn@#1@PL\endcsname%
	\fi%
	\else%
	\AC@acf{#1}s%
	\fi%
}%
\def\Acsp#1{\AC@uppercase@firsttrue\acsp{#1}\AC@uppercase@firstfalse}%
\def\AC@acsp#1{%
	\ifcsname fn@#1@PL\endcsname%
	\ifAC@uppercase@first%
	\expandafter\expandafter\expandafter\MakeUppercase\csname fn@#1@PL\endcsname%
	\else%
	\csname fn@#1@PL\endcsname%
	\fi%
	\else%
	\AC@acs{#1}s%
	\fi%
}%
\edef\AC@uppercase@write{\string\ifAC@uppercase@first\string\expandafter\string\MakeUppercase\string\fi\space}%
\def\AC@acrodef#1[#2]#3{%
	\@bsphack%
	\protected@write\@auxout{}{%
		\string\newacro{#1}[#2]{\AC@uppercase@write #3}%
	}\@esphack%
}%
\def\Acl#1{\AC@uppercase@firsttrue\acl{#1}\AC@uppercase@firstfalse}
\def\Acf#1{\AC@uppercase@firsttrue\acf{#1}\AC@uppercase@firstfalse}
\def\Ac#1{\AC@uppercase@firsttrue\ac{#1}\AC@uppercase@firstfalse}
\def\Acs#1{\AC@uppercase@firsttrue\acs{#1}\AC@uppercase@firstfalse}

\newtheorem{theorem}{Theorem}

\acrodef{SIC}{successive interference cancellation}
\acrodef{PAPR}{peak-to-average-power ratio}
\acrodef{APAC}{aperiodic autocorrelation}
\acrodef{OFDM}{orthogonal frequency division multiplexing}
\acrodef{DFT}{discrete Fourier transform}
\acrodef{DC}{direct current}
\acrodef{CS}{complementary sequence}
\acrodef{GCP}{Golay complementary pair}
\acrodef{ANF}{algebraic normal form}
\acrodef{PSK}{phase-shift keying}
\acrodef{QAM}{quadrature amplitude modulation}
\acrodef{QPSK}{quadrature phase shift keying}
\acrodef{GDJ}{Golay-Davis-Jedwab}
\acrodef{PMEPR}{peak-to-mean envelope power ratios}
\acrodef{FFT}{fast Fourier transform}
\acrodef{BER}{bit-error rate}
\acrodef{SNR}{signal-to-noise ratio}
\acrodef{4G}{Fourth Generation}
\acrodef{5G}{Fifth Generation}
\acrodef{NR}{5G New Radio}
\acrodef{LTE}{Long-Term Evolution}
\acrodef{PTS}{partial transmit sequences}
\acrodef{PSD}{power spectral density}
\acrodef{LDPC}{low-density parity check}
\acrodef{SE}{spectral efficiency}
\acrodef{eLAA}{enhanced licensed-assisted access}
\acrodef{NR-U}{NR-Unlicensed}
\acrodef{RM}{Reed-Muller}
\acrodef{AE}{autoencoder}
\acrodef{DNN}{deep neural network}
\acrodef{OFDM-AE}{OFDM-based autoencoder}
\acrodef{DL}{deep learning}
\acrodef{CP}{cyclic prefix}
\acrodef{AWGN}{additive white Gaussian noise}
\acrodef{P2C}{polar-to-Cartesian}
\acrodef{CFR}{channel frequency response}
\acrodef{ReLU}{rectified linear unit}
\acrodef{MMSE}{minimum mean square error}
\acrodef{BPSK}{binary phase shift keying}
\acrodef{BLER}{block error rate}
\acrodef{ML}{machine learning}
\acrodef{PHY}{physical layer}
\acrodef{PA}{power amplifier}
\begin{document}
\title{ 
Golay Layer: Limiting Peak-to-Average Power Ratio for OFDM-based Autoencoders
}
\author{Alphan~\c{S}ahin,~\IEEEmembership{Member,~IEEE} and David~W.~Matolak,~\IEEEmembership{Member,~IEEE} 
\thanks{Alphan~\c{S}ahin and David W. Matolak are with the University of South Carolina, Columbia, SC. E-mail: asahin@mailbox.sc.edu, matolak@cec.sc.edu
}
}
\maketitle

\begin{abstract}
In this study, we propose a differentiable layer for  \acp{OFDM-AE}  to avoid high instantaneous power without regularizing the cost function used during the training. 
The proposed approach relies on the manipulation of the parameters of a set of  functions that yield  \acp{CS} through a \ac{DNN}. 
We guarantee the \ac{PAPR} of each \ac{OFDM-AE} symbol to be less than or equal to 3 dB.
We also show how to normalize the mean power by using the functions in addition to \ac{PAPR}. 
The introduced layer admits auxiliary parameters that allow one to control the amplitude and phase deviations in the frequency domain. 
Numerical results show that \acp{DNN} at the transmitter and receiver can  achieve reliable communications under this protection layer at the expense of complexity.
\end{abstract}

\acresetall

\section{Introduction}
 
Historically, the \ac{PHY} of a communication system has been designed based on well-engineered signal processing blocks. 
This design philosophy has recently been disrupted with the success of \ac{ML} over handcrafted designs in various fields. In \cite{Shea_2016} and  \cite{Shea_2017}, the composite behavior of the transmitter-channel-receiver is represented as an \ac{AE} where the signal processing blocks at the transmitter and receiver are replaced with neural networks. 
It has been shown that the \ac{AE} can outperform a Hamming-coded \ac{BPSK} scheme and achieves end-to-end learning for small-scale problems without prescribing specific signal processing blocks.
However, when a neural network is applied to a large scale \ac{PHY} design, the training complexity, reliability, scalability, and unexplainable nature of trained neural networks become major issues. 
Therefore, new methods that can facilitate the \ac{PHY} automation need to be developed based on the constraints and requirements in the communication systems. 

To this end, one direction is to blend the tools that we have already employed in the communication systems with the machine learning blocks. For example, in \cite{Felix_2018},  an \ac{AE} was combined with \ac{OFDM}, called \ac{OFDM-AE}, to reduce the complexity of synchronization stages for a single-carrier modulation scheme in \cite{Dorner_2018}. \ac{OFDM-AE} is appealing because \ac{OFDM} is one of the most used schemes in today's wireless communication standards such as 3GPP \ac{LTE}, \ac{NR}, and IEEE 802.11 Wi-Fi. It can be implemented via relatively low-complexity architectures while   accommodating  many crucial aspects of communications, e.g., user multiplexing, multiple antennas, and simple equalization. In addition, it can easily multiplex different type waveforms, including the ones based on \acp{OFDM-AE}, in the frequency domain. In the literature, \acp{OFDM-AE} have currently been investigated under  multiple antennas \cite{Ji_2019}, training under an unknown channel model \cite{Aoudia_2019}, one-bit quantization \cite{Balevi_2019}, and \ac{PAPR} mitigation \cite{kim_2018,  Miao_2019}. In this study, we also focus on \ac{OFDM-AE} and address the issue of designing \ac{OFDM-AE} with low \ac{PAPR}.

The methods that overcome the hardware non-linearity for \ac{OFDM-AE} in the literature may be grouped into two main categories. In the first category, the approaches rely on the fact that \acp{AE} are able to compensate for multiple effects jointly. For example, in \cite{Felix_2018}, the AM-AM distortion due to a \ac{PA} is modeled as a third-order non-linear function with a normalized input and the model is included as part of the channel. Similarly, a coarse quantization is considered as a distortion in the channel in \cite{Balevi_2019}. The main challenge in this category is either the availability of an accurate differentiable non-linearity model or the price paid for more comprehensive training as in \cite{Aoudia_2019}. The methods in the second category 
aim at \ac{AE} design to combat the non-linearity. One common approach in the literature is to regularize the cost function used during the training. For example, in \cite{kim_2018} and \cite{Ji_2019}, the cost function is a summation of a metric related to \ac{PAPR} and another metric for \ac{BER}, which are similar to the techniques used for traditional \ac{OFDM} \cite{Tom_2016}. In \cite{Miao_2019}, both \ac{DFT}-spread OFDM and a regularization term are considered to reduce the  \ac{PAPR} for light communications. However, these methods require more sophisticated training/optimization procedures, which may increase the training complexity in practice. In this study, we address this issue and propose a protection layer, called Golay layer,  by exploiting the properties of \acp{CS} to limit instantaneous power fluctuations without introducing a regularization term to the cost function.

The \acp{CS} were introduced by M. Golay in 1961 \cite{Golay_1961}. The \ac{PAPR} of an \ac{OFDM} symbol generated with a \ac{CS}  is less than or equal to 3 dB  \cite{boyd_1986,Popovic_1991}. 
In \cite{davis_1999}, it has been shown that a specific \ac{RM} code along with \ac{PSK} constellation results in \acp{CS}. Hence, this particular scheme
achieves coding gain while ensuring 3 dB \ac{PAPR}  for \ac{OFDM} symbols. 
In the literature, there are numerous studies to construct distinct \acp{CS}. The reader is referred to \cite{Budisin_1990} and \cite{Budisin_1990_ml} for iterative methods, \cite{robing_2001} and \cite{Li_2010} for the offset method, \cite{budisin_2018} for a method based on Gaussian integers, and \cite{davis_1999} and \cite{Sahin_2018} for algebraic approaches for the \ac{CS} synthesis. A survey related to \acp{CS} can also be found in \cite{parker_2003}.

Our main contribution in this study is the derivation of a layer that converts a set of real numbers to another set of real numbers based on the framework given in \cite{Sahin_2018} to guarantee maximum 3 dB \ac{PAPR} for an \ac{OFDM-AE}. We show that the proposed layer is compatible with  the existing layers in \ac{ML} literature  and the training methods (e.g., backpropagation) as the corresponding gradients can be expressed in closed-form, and the amplitude and the phase of the elements of synthesized \ac{CS}, which can be tuned through neural networks independently without affecting the \ac{PAPR}, are real numbers. With  numerical results, we show that it is possible to achieve power-efficient reliable \ac{OFDM-AE} with the proposed layer.


The rest of the paper is organized as follows. In Section \ref{sec:prelim}, we provide preliminary discussions on sequences. In Section \ref{sec:Glayer}, we introduce the Golay layer. In Section \ref{sec:numerical}, we present numerical results and compare it with \ac{OFDM} with polar code. We conclude the paper in Section \ref{sec:conclusion}.

{\em Notation:} The sets of complex numbers, real numbers, non-negative real numbers, integers,  non-negative integers,  positive integers, and integers modulo $\numberOfPointsForPSK$ are denoted by $\complexNumbers$,  $\realNumbers$, $\realsNonnegative$,  $\integers$, $\integersNonnegative$, $\integersPositive$, and $\integers_\numberOfPointsForPSK$, respectively. The set of $\numberOfIterations$-dimensional integers where each element is in $\integers_\numberOfPointsForPSK$  is denoted by $\integers^\numberOfIterations_\numberOfPointsForPSK$.  
The modulo 2 operation is denoted by $(\cdot)_2$.
 The constant $\constantj$ denotes $\sqrt{-1}$. 

\section{Preliminaries}
\label{sec:prelim}

An \ac{OFDM} symbol with the symbol duration $\symbolDuration$ can be expressed in continuous time as a polynomial given by
\begin{align}
\polySeq[\seqGa][\polyVariable] \triangleq \eleGa[\lengthGaGb-1]\polyVariable^{\lengthGaGb-1} + \eleGa[\lengthGaGb-2]\polyVariable^{\lengthGaGb-2}+ \dots + \eleGa[0]~,
\label{eq:polySeq}
\end{align}
where $\lengthGaGb$ is the number of subcarriers, $\seqGa$ is a sequence of length $\lengthGaGb$, and $\polyVariable\in\{\constante^{\constantj\frac{2\pi\timeVar}{\symbolDuration}}| 0\le\timeVar <\symbolDuration
\}$. The properties of the sequence $\seqGa$ play a central role for the link-level performance of an \ac{OFDM}-based communication system. This is due the fact that the sequence $\seqGa$ can be not only a representation of $\numberOfInfoBits$ information bits over $\complexNumbers^\lengthGaGb$ from the perspective of modulation and coding but also  a key component that determines the waveform characteristics in time and frequency. For example, the instantaneous envelope power of the \ac{OFDM} symbol can be written as
\begin{align}
|\polySeq[\seqGa][\polyVariable]|^2 &
\bigg\rvert_{\polyVariable=\constante^{\constantj\frac{2\pi\timeVar}{\symbolDuration}}} 
= \sum_{\lagForCorrelation=-\lengthGaGb+1}^{\lengthGaGb-1}\apac[\seqGa][\lagForCorrelation]\constante^{\constantj\frac{2\pi\timeVar}{\symbolDuration}\lagForCorrelation}
~,
\label{eq:instantaneousPower}
\end{align} 
where $\apac[\seqGa][\lagForCorrelation]$ is the \ac{APAC} of the sequence $\seqGa$ at the $\lagForCorrelation$th lag \cite{Sahin_2018}. To minimize  instantaneous power fluctuations, by inferring \eqref{eq:instantaneousPower},  the sequence $\seqGa$ should have good \ac{APAC} properties, i.e., low $|\apac[\seqGa][\lagForCorrelation]|$ for $\lagForCorrelation\neq0$.

\subsection{Complementary Sequences}
\label{subsec:CS}

The sequence pair  $(\seqGa,\seqGb)$ of length  $\lengthGaGb$ is called a \ac{GCP} if $\apac[\seqGa][\lagForCorrelation]+\apac[\seqGb][\lagForCorrelation] = 0$ for $\lagForCorrelation~\neq0~$ \cite{Golay_1961}. Each sequence in a \ac{GCP} is called a \ac{CS}. Equivalently, the GCP $(\seqGa,\seqGb)$ can be defined as \cite{parker_2003}
\begin{align}
|\polySeq[\seqGa][\polyVariable]|^2+|\polySeq[\seqGb][\polyVariable]|^2\bigg\rvert_{\polyVariable=\constante^{\constantj\frac{2\pi\timeVar}{\symbolDuration}}} =\underbrace{\apac[\seqGa][0]+\apac[\seqGb][0]}_{\text{constant}}~.
\label{eq:timeDomainGCP}
\end{align}
In other words, the sum of the instantaneous power of two \ac{OFDM} symbols generated with $\seqGa$ and $\seqGb$  adds up to a constant if $\seqGa$ and $\seqGb$ form a \ac{GCP}. Hence, $|\polySeq[\seqGa][{
	\constante^{\constantj\frac{2\pi\timeVar}{\symbolDuration}}
}]|^2$ is always bounded, i.e.,
$\max_{\timeVar}|\polySeq[\seqGa][{
	\constante^{\constantj\frac{2\pi\timeVar}{\symbolDuration}}
}]|^2 \le \apac[\seqGa][0]+\apac[\seqGb][0]$. As a result, the \ac{PAPR} of the \ac{OFDM} symbol $\polySeq[\seqGa][{
\constante^{\constantj\frac{2\pi\timeVar}{\symbolDuration}}
}]$  is  less than or equal to $10\log_{10}(2)\approx3$~dB if $\apac[\seqGa][0]=\apac[\seqGb][0]$ \cite{boyd_1986,Popovic_1991}.

\subsection{Representation of a Sequence}
\label{subsec:algebraic_sequence}
Let $\funcfForANF$ be a function that maps from $\integers^\numberOfIterations_2=\{(\monomial[1],\monomial[2],\dots, \monomial[\numberOfIterations])| \monomial[\indexFirstOrderMonomial]\in\integers_2\}$ to $\generalSet$ as $\funcfForANF:\integers^\numberOfIterations_2\rightarrow\generalSet$, where $\generalSet$ is a an arbitrary set. In this study, a sequence $\seqf$ of length $2^\numberOfIterations$ is composed by listing the values of the function $\funcfForANF(\monomial[1],\monomial[2],\dots, \monomial[\numberOfIterations])$  as $(\monomial[1],\monomial[2],\dots, \monomial[\numberOfIterations])$ ranges over its $2^\numberOfIterations$ values in lexicographic order. In other words, the $(\varMonomial +1)$th element of the sequence $\seqf$ is equal to $\funcfForANF(\monomial[1],\monomial[2],\dots, \monomial[\numberOfIterations])$ where $\varMonomial = \sum_{\indexFirstOrderMonomial=1}^{\numberOfIterations}\monomial[\indexFirstOrderMonomial]2^{\numberOfIterations-\indexFirstOrderMonomial}$  (i.e., the most significant bit is $\monomial[1]$). 
For the sake of simplifying the notation, we denote the sequence $(\monomial[1],\monomial[2],\dots, \monomial[\numberOfIterations])$ and the function $\funcfForANF(\monomial[1],\monomial[2],\dots, \monomial[\numberOfIterations])$ as $\seqx$ and  $\funcfForANF(\seqx)$, respectively. 

The function $\funcfForANF(\seqx)$ is called a generalized Boolean function for  $\generalSet=\integers_\numberOfPointsForPSK$ for $\numberOfPointsForPSK\in\integersNonnegative$. If $\numberOfPointsForPSK=2$, $\funcfForANF(\seqx)$ is a Boolean function. The function $\funcfForANF(\seqx)$ can be uniquely expressed as a linear combination of the monomials over $\generalSet$  as
\begin{align}
\hspace{-2mm}\funcfForANF(\seqx)= \sum_{\indexMonomial=0}^{2^\numberOfIterations-1} \coeffientsANF[\indexMonomial]\prod_{\indexFirstOrderMonomial=1}^{\numberOfIterations} \monomial[\indexFirstOrderMonomial]^{\orderMonomial[\indexFirstOrderMonomial]} = \coeffientsANF[0]1+ \dots+ \coeffientsANF[2^\numberOfIterations-1]\monomial[1]\monomial[2]\mydots\monomial[\numberOfIterations]~,
\label{eq:ANF}
\end{align} 
where $\coeffientsANF[\indexMonomial]\in \generalSet$, $\indexMonomial = \sum_{\indexFirstOrderMonomial=1}^{\numberOfIterations}\orderMonomial[\indexFirstOrderMonomial]2^{\numberOfIterations-\indexFirstOrderMonomial}$ for $\orderMonomial[\indexFirstOrderMonomial]\in\integers_2$. If $\funcfForANF(\seqx)$ is over $\generalSet=\realNumbers$, each monomial coefficient belongs to $\realNumbers$, i.e.,  $\coeffientsANF[\indexMonomial]\in \realNumbers$. The monomials construct a vector space over $\realNumbers$, where its dimension is $2^\numberOfIterations$. Therefore, different sets of $\{\coeffientsANF[\indexMonomial] |\indexMonomial=0,\mydots,2^{\numberOfIterations}-1\}$ lead to different sequences. This is also true for $\generalSet=\integers_\numberOfPointsForPSK$ as the monomials are linearly independent.

\section{Golay Layer}
\label{sec:Glayer}

A direct \ac{GCP} construction based on four basic functions 
is given in \cite{Sahin_2018}. By focusing only on one of the sequences in a \ac{GCP}, we can restate the theorem in \cite{Sahin_2018} as follows:

\begin{theorem}[\ac{CS} Construction]
	\label{th:reduced}
	Let $(\seqGa,\seqGb)$ be a \ac{GCP} of length $\lengthGaGb$ and $\seqPermutationCompShift=(\permutationMono[\indexIteration])_{\indexIteration=1}^{\numberOfIterations}$ be a sequence defined by a permutation of $\{1,2,\dots,\numberOfIterations\}$. For any $\deviationAmplitude,\deviationPhase \in\realsNonnegative$, $\separationGolay[\indexIteration]\in \integersNonnegative$, $\scaleEexp[\indexIteration]\in\realNumbers$, and $\angleexpAll[\indexIteration]\in\realsNonnegative$ for $\indexIteration=0,1,\mydots,\numberOfIterations$, let
	\begin{align}
	&\funcfForFinalAmplitude(\seqx)
	=\deviationAmplitude\left(\scaleEexp[\numberOfIterations]\monomial[{\permutationMono[{\numberOfIterations}]}]+{ \sum_{\indexIteration=1}^{\numberOfIterations-1}\scaleEexp[\indexIteration](\monomial[{\permutationMono[{\indexIteration}]}] +\monomial[{\permutationMono[{\indexIteration+1}]}])_2 }+\arbitraryScaleE\right)\label{eq:realPartReduced}~,
	\\	
	&\funcfForFinalPhase(\seqx)
	= \pi\left({\sum_{\indexIteration=1}^{\numberOfIterations-1}\monomial[{\permutationMono[{\indexIteration}]}]\monomial[{\permutationMono[{\indexIteration+1}]}]}\right)+\deviationPhase\left(\sum_{\indexIteration=1}^\numberOfIterations \angleexpAll[\indexIteration]\monomial[{\permutationMono[{\indexIteration}]}]+  \arbitraryPhaseK\right)\label{eq:imagPartReduced}~,
	\\	
	&\funcfForCommonShift(\seqx) =\sum_{\indexIteration=1}^\numberOfIterations\separationGolay[\indexIteration]\monomial[{\permutationMono[{\indexIteration}]}]~,
	\label{eq:shift}
	\\
	&\funcfForCommonOrder(\seqx,\polyVariable)=\polySeq[{\seqGa}][\polyVariable](1-\monomial[{\permutationMono[{1}]}])_2+\polySeq[{\seqGb}][\polyVariable]\monomial[{\permutationMono[{1}]}]~.\label{eq:order}
	\end{align}
	Then, the sequence $\seqGc$ where its polynomial representation is given by
	\begin{align}
	\polySeq[{\seqGc}][\polyVariable] &= 
	\sum_{\varMonomial=0}^{2^\numberOfIterations-1} 
	\funcfForCommonOrder(\seqx,\polyVariable)\times
	\constante^{{\funcfForFinalAmplitude(\seqx)   +\constantj \funcfForFinalPhase(\seqx)}}
	\times
	\polyVariable^{\funcfForCommonShift(\seqx) + \varMonomial\lengthGaGb}\label{eq:encodedFOFDMonly}
	\end{align}
	is a \ac{CS} of length $\lengthGaGb2^\numberOfIterations+\sum_{\indexIteration=1}^\numberOfIterations\separationGolay[\indexIteration]$.
\end{theorem}

The polynomial given in \eqref{eq:encodedFOFDMonly} corresponds to an \ac{OFDM} symbol  where the sequence in the frequency domain is a \ac{CS} $\seqGc$. Therefore, based on the discussions in Section~\ref{subsec:CS}, the instantaneous power for the corresponding \ac{OFDM} symbol  is bounded. 
In general, the sequence $\seqGc$ is a function  of an initial \ac{GCP} $(\seqGa,\seqGb)$, $\deviationAmplitude$, $\deviationPhase$, $\seqPermutationCompShift$, $\separationGolay[\indexIteration]$, $\scaleEexp[\indexIteration]$, and $\angleexpAll[\indexIteration]$ for $\indexIteration=0,1,\mydots,\numberOfIterations$,  and formed by the functions given in \eqref{eq:realPartReduced}-\eqref{eq:order}.
 While the functions related to the real and imaginary part of the exponent, i.e. $\funcfForFinalAmplitude(\seqx)$ and $\funcfForFinalPhase(\seqx)$, allow one to adjust the amplitude and phase of the elements of \acp{CS}, respectively, $\funcfForCommonShift(\seqx)$ alters the locations of the initial sequences, i.e., $\seqGa$ and $\seqGb$, in the frequency domain. The function $\funcfForCommonOrder(\seqx,\polyVariable)$ determines which of the initial sequences is modified through $\funcfForFinalAmplitude(\seqx)$, $\funcfForFinalPhase(\seqx)$, and $\funcfForCommonShift(\seqx)$ based on \eqref{eq:encodedFOFDMonly}. The parameters $\deviationAmplitude$ and $\deviationPhase$ are the auxiliary variables introduced in this study to control the amount of the amplitude and phase deviations. For example, if $\deviationAmplitude=0$, the parameters related to the amplitude, i.e., $\scaleEexp[\indexIteration]$ for $\indexIteration=0,1,\mydots,\numberOfIterations$, do not determine the final sequence. Similarly, if $\deviationPhase=0$, the synthesized sequence is a not a function of $\angleexpAll[\indexIteration]$ for $\indexIteration=0,1,\mydots,\numberOfIterations$.

In the literature, there are many studies on the enumeration of distinct \acp{CS} such that their elements are in the traditional constellations such as $M$-\ac{QAM}, particularly after the discovery of the connection between \ac{RM} codes and \acp{CS} in \cite{davis_1999}. In their pioneering  work,  Davis and Jedwab showed that the \acp{CS} can occur as the elements of the cosets of the first-order \ac{RM} code within the second-order \ac{RM} code where the elements of  the  \ac{CS} are in \ac{PSK} constellation. The corresponding function that defines the \ac{RM} code can be seen in \eqref{eq:imagPartReduced}. For instance, by choosing $\angleexpAll[\indexIteration] \in \integers_{2^q}$ for $\indexIteration=0,1,\mydots,\numberOfIterations$ and $q\in\integersPositive$, $\deviationPhase=2\pi/{2^q}$, $\deviationAmplitude=0$, $\funcfForCommonOrder(\seqx,\polyVariable)=1$, \eqref{eq:encodedFOFDMonly} yields a coded \ac{OFDM} symbol with the maximum of $3$ dB \ac{PAPR}, where the \ac{CS} in the frequency domain is a codeword from  the coset of the first-order \ac{RM} code over $\integers_{2^q}$ with $2^q$-\ac{PSK} modulation. 
In \cite{Sahin_2018}, several rules are introduced for \ac{CS} with $M$-\ac{QAM} constellation for Theorem~\ref{th:reduced}. We refer the reader to other approaches based on Gaussian integers and offset methods in \cite{Li_2010} and \cite{budisin_2018}, respectively.
Nevertheless, \ac{CS}-based encoding and decoding are still challenging tasks. For example, it is not trivial to map the information bits to the parameters introduced in Theorem~\ref{th:reduced}. In addition, to the best of our knowledge, there is no study that shows that $M$-\ac{QAM} is the optimum constellation to achieve good \ac{BER} performance for an encoder that generates \acp{CS}. The decoding at the receiver side can be nontrivial as the modulation and coding are not independent operations in Theorem~\ref{th:reduced} (e.g., see the discussions in \cite{davis_1999}). These design issues motivate us to investigate \ac{CS}-based encoding and decoding under \acp{OFDM-AE}.


The key observation that we exploit in this study is that the variables in Theorem~\ref{th:reduced} can be tuned with a \ac{DNN} based on the information bits without affecting the \ac{PAPR}. From the perspective of \ac{ML}-based PHY design,  this approach can also be considered as a framework for developing new layers for power-efficient \ac{OFDM-AE} as Theorem~\ref{th:reduced} limits the \ac{PAPR} to be less than 3 dB {\em without} introducing constraints on the cost function used during the training while being compatible with state-of-the-art \ac{ML} approaches. For example, the parameters which control the amplitude and phase of the elements of the \ac{CS}, i.e., $\scaleEexp[\indexIteration]$, and $\angleexpAll[\indexIteration]$, are real numbers and inherently determine the complex numbers in polar coordinate. The gradients for the phase and amplitude functions  can also be calculated in closed-form expressions as the functions given in \eqref{eq:realPartReduced} and \eqref{eq:imagPartReduced} are multivariate polynomials. Therefore, the backpropagation can be used without any restriction. 

%
%
%

\subsection{Basic Golay Layer}

\def\totalPowerScale{\gamma}

Without loss of generality, in this study, we assume that $\seqGa=(1)$ and $\seqGb=(1)$, the values of $\separationGolay[\indexIteration]$ and $\seqPermutationCompShift$ are controlled either by a communication network or they are prescribed to adjust the position of the non-zero elements of the encoded \ac{CS}. We define the basic Golay layer as 
\begin{align}
y=f(\seqx)={\funcfForFinalAmplitude(\seqx)   +\constantj \funcfForFinalPhase(\seqx)}.
\end{align}
Based on \eqref{eq:imagPartReduced}, the derivative of $\funcfForFinalPhase(\seqx)$ with respect to $\angleexpAll[\indexIteration]$ and $\arbitraryPhaseK$, and the derivative of $\funcgForFinalPhase(\seqx)$ with respect to  $\angleexpAll[\indexIteration]$  can be calculated as
\begin{align}
\frac{\partial \funcfForFinalPhase(\seqx)}{\partial \angleexpAll[\indexIteration] } = \deviationPhase \monomial[{\permutationMono[{\indexIteration}]}]~,
\end{align}
and
\begin{align}
\frac{\partial \funcfForFinalPhase(\seqx)}{\partial \arbitraryPhaseK }= \deviationPhase~,
\end{align}
respectively. 


The mean \ac{OFDM} symbol power is a function of the amplitude of each element of the \ac{CS}. However, Theorem~\ref{th:reduced} without any constraint does not guarantee a fixed mean power for each set of $\scaleEexp[\indexIteration]$ for $\indexIteration = 0,1,\mydots,\numberOfIterations$. To resolve this issue, we choose $\arbitraryScaleE$ as a normalization parameter  since it is a constant term in \eqref{eq:realPartReduced} (i.e, it scales all the elements of the synthesized \ac{CS} in \eqref{eq:encodedFOFDMonly}), and introduce the condition given by
\begin{align}
\deviationAmplitude\arbitraryScaleE = \frac{1}{2}\sum_{\indexIteration=1}^{\numberOfIterations}\ln{\frac{1+\constante^{2\deviationAmplitude\scaleEexp[\indexIteration]}}{2}}.
\label{eq:normalizationCoef}
\end{align}
The condition in \eqref{eq:normalizationCoef} can be derived as follows: The parameter $\scaleEexp[\indexIteration]$ scales half of the elements of the \ac{CS} by $\constante^{\deviationAmplitude\scaleEexp[\indexIteration]}$ due to the monomials in \eqref{eq:realPartReduced}. Therefore, the \ac{CS} power is scaled by ${(1+\constante^{2\deviationAmplitude\scaleEexp[\indexIteration]}})/{2}$. For all $\scaleEexp[\indexIteration|\indexIteration = 1,2,\mydots,\numberOfIterations]$, the total  factor can be calculated as $\totalPowerScale=\prod_{\indexIteration=1}^{\numberOfIterations}{(1+\constante^{2\deviationAmplitude\scaleEexp[\indexIteration]}})/{2}$. Hence, to normalize the power, $\constante^{2\deviationAmplitude\scaleEexp[\indexIteration]}=1/\totalPowerScale$, which results in \eqref{eq:normalizationCoef}. Under the condition  in \eqref{eq:normalizationCoef}, the derivative of $\funcfForFinalAmplitude(\seqx)$ with respect to $\scaleEexp[\indexIteration]$ for $\indexIteration = 1,2,\mydots,\numberOfIterations$ can be calculated as
\begin{align}
\frac{\partial \funcfForFinalPhase(\seqx)}{\partial \scaleEexp[\indexIteration] } = \begin{cases} 
\deviationAmplitude(\monomial[{\permutationMono[{\indexIteration}]}] +\monomial[{\permutationMono[{\indexIteration+1}]}])_2 - \frac{\constante^{2\deviationAmplitude\scaleEexp[\indexIteration]}}{1+\constante^{2\deviationAmplitude\scaleEexp[\indexIteration]}} & \indexIteration<\numberOfIterations \\
\deviationAmplitude(\monomial[{\permutationMono[{\numberOfIterations}]}])_2 - \frac{\constante^{2\deviationAmplitude\scaleEexp[\numberOfIterations]}}{1+\constante^{2\deviationAmplitude\scaleEexp[\numberOfIterations]}} & \indexIteration=\numberOfIterations
\end{cases}
\end{align}
as 
\begin{align}
\frac{\partial \arbitraryScaleE}{\partial \scaleEexp[\indexIteration] }= - \frac{\constante^{2\deviationAmplitude\scaleEexp[\numberOfIterations]}}{1+\constante^{2\deviationAmplitude\scaleEexp[\numberOfIterations]}}~.
\end{align}
As a result, the basic Golay layer consists of two differentiable functions with $\numberOfIterations$ inputs for amplitude and $\numberOfIterations+1$ inputs for the phase and returns $2^{m}$ complex  (or $2^{\numberOfIterations+1}$ real) values as output.
Let $\setMiniBatch$ be a minibatch of size $\sizeofBatch$. The derivative of the loss $\lossFunction$ (e.g., cross-entropy function) with respect to $\indexSampleInBatch$th  $\angleexpAllBatch[\indexIteration][\indexSampleInBatch]$ for $\indexIteration = 0,1,\mydots,\numberOfIterations$ and   $\scaleEexpBatch[\indexIteration][\indexSampleInBatch]$ for $\indexIteration = 1,2,\mydots,\numberOfIterations$ can then be obtained as
\begin{align}
\frac{\partial \lossFunction}{\partial \angleexpAllBatch[\indexIteration][\indexSampleInBatch] } =\sum_{\varMonomial=0}^{2^\numberOfIterations-1} \frac{\partial \funcfForFinalPhase(\seqx)}{\partial \angleexpAllBatch[\indexIteration][\indexSampleInBatch] } \frac{\partial \lossFunction}{\partial \funcfForFinalPhase(\seqx)}~,
\\
\frac{\partial \lossFunction}{\partial \scaleEexpBatch[\indexIteration][\indexSampleInBatch] } =\sum_{\varMonomial=0}^{2^\numberOfIterations-1} \frac{\partial \funcfForFinalAmplitude(\seqx)}{\partial \scaleEexpBatch[\indexIteration][\indexSampleInBatch] } \frac{\partial \lossFunction}{\partial \funcfForFinalAmplitude(\seqx)}~.
\end{align}
where $\indexSampleInBatch$ denotes the  $\indexSampleInBatch$th sample in the minibatch $\setMiniBatch$, and $\varMonomial$ and $\seqx$ are defined in Section \ref{subsec:algebraic_sequence}.

\begin{figure*}[t]
	\centering
	\subfloat[Transmitter and receiver block diagrams for OFDM-AE with Golay layer.]{\includegraphics[width =5.6in]{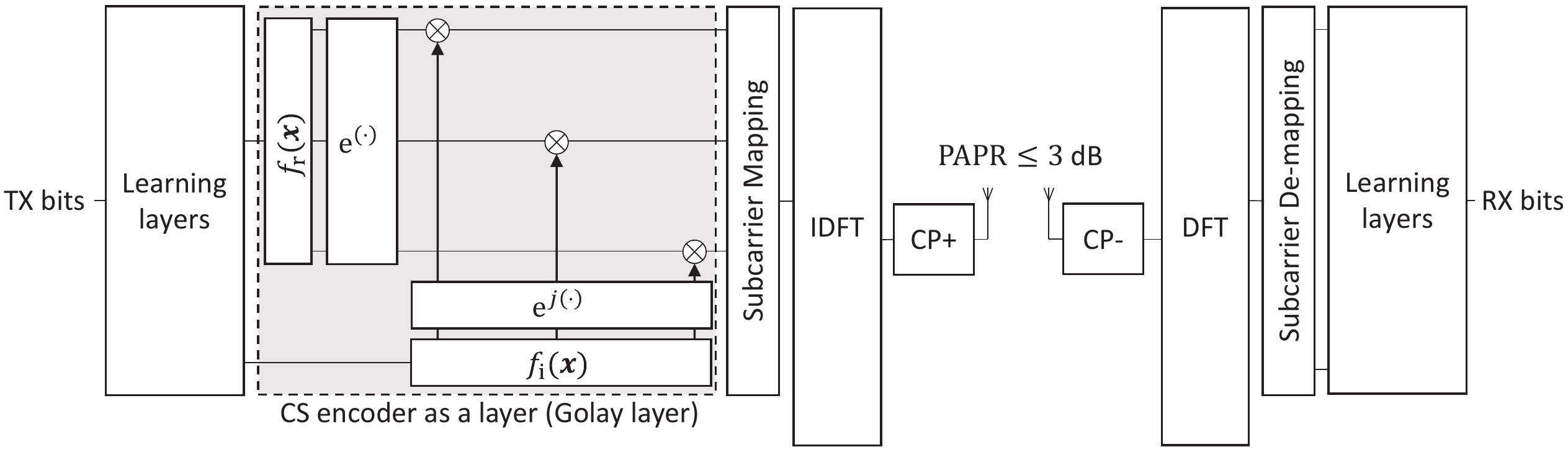}
		\label{fig:txrx}}
	\\
	\subfloat[An \ac{AE} representation of the complete link.]{\includegraphics[width =7in]{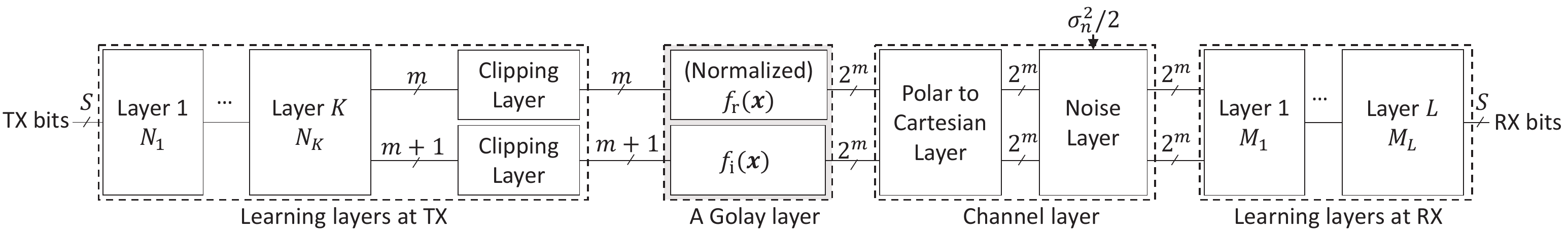}
		\label{fig:autoencoder}}
	\caption{\ac{OFDM-AE} with Golay layer.}
	\label{fig:txrxae}
\end{figure*}

\def\indexLayerTX{k}
\def\indexLayerRX{l}
\def\numberOfLayersAtTX{K}
\def\numberOfLayersAtRX{L}
\def\numberOfOutputNodeAtTX[#1]{N_{#1}}
\def\numberOfOutputNodeAtRX[#1]{M_{#1}}

In \figurename~\ref{fig:txrxae}\subref{fig:txrx}, we provide the transmitter and receiver block diagrams for OFDM-AE with the Golay layer. At the transmitter, we consider a \ac{DNN} with $\numberOfLayersAtTX$ layers where the $\indexLayerTX$th layer has $\numberOfOutputNodeAtTX[{\indexLayerTX}]$ output nodes. The \ac{DNN} at the transmitter maps a sequence of information bits to the aforementioned $2\numberOfIterations+1$ parameters and learns the bit mapping. The Golay layer calculates the amplitude and the phase parameters of the corresponding \ac{CS} based on the output of the preceding layer. After mapping the synthesized \ac{CS} to the \ac{OFDM} subcarriers, the inverse \ac{DFT} of the mapped \ac{CS} is calculated. A \ac{CP} can also be prepended to the generated symbol before transmitting the \ac{OFDM} symbol. At the receiver side, the \ac{CP} is removed and the \ac{DFT} of the received signal is calculated. After the subcarrier de-mapping, another \ac{DNN} with $\numberOfLayersAtRX$ layers where the $\indexLayerRX$th layer has $\numberOfOutputNodeAtRX[{\indexLayerRX}]$ output nodes at the receiver side obtain the information bits from the received \ac{CS}.

\subsection{Supporting Layers}
\def\minClip{r_{\rm min}}
\def\maxClip{r_{\rm max}}
In this study, we do not introduce any restriction for the \acp{DNN} at the transmitter and receiver.  However, based on our computer trials, several supporting layers can be helpful for training and manipulating the parameters of the Golay layer in a controlled manner. For example, a clipping layer that limits the range of the parameters for the Golay layer can avoid the growth of exponent in \eqref{eq:encodedFOFDMonly}. The clipping layer which allows the variable between $\minClip$ and $\maxClip$ to pass without any distortion can be defined as
\begin{align}
y=f(x)=\min\{\max\{x,\minClip\},\maxClip\}
\end{align}
where $\minClip\le\maxClip$. The derivative of the loss $\lossFunction$ with respect to the variable $x$ can be calculated as $\partial\lossFunction/\partial y = \partial\lossFunction/\partial x$ for $\minClip\le x \le\maxClip$, otherwise it is zero. 

\def\channelCoefficient{h}
\def\noiseVariance{\sigma_{\rm n}^2}
Another auxiliary function which may be needed during the training is the \ac{P2C} layer which can be defined as
\begin{align}
a =&f(x,y)=\Re\{\constante^{x+\constantj y}\} = \constante^{x}\cos(y)~,\\
b =&f(x,y)=\Im\{\constante^{x+\constantj y}\} = \constante^{x}\sin(y)~.
\end{align}
The derivative of the loss with respect to the variable $x$ and $y$ are
$\partial\lossFunction/\partial x = \partial\lossFunction/\partial a \times\constante^{x}\cos(y) + \partial\lossFunction/\partial b\times \constante^{x}\sin(y)$ and $\partial\lossFunction/\partial y = -\partial\lossFunction/\partial a\times \constante^{x}\sin(y)+\partial\lossFunction/\partial b\times\constante^{x}\cos(y)$, respectively. 
We also consider an \ac{AWGN} layer which  adds noise to the real and imaginary components with the variance $\noiseVariance/2$ after the \ac{P2C} layer for the sake of training the \ac{OFDM-AE} offline under certain \ac{SNR}. 

A complete \ac{AE} representation of the link with the supporting layers is shown in \figurename~\ref{fig:txrxae}\subref{fig:autoencoder}. First, the $\numberOfInfoBits$ information bits are processed by a \ac{DNN} with $\numberOfLayersAtTX$ layers. The  $2 \numberOfIterations+1 $ outputs of the $\numberOfLayersAtTX$th layer at the transmitter are then limited by the two parallel clipping layers. After the calculation of $\funcfForFinalAmplitude(\seqx)$ and $\funcfForFinalPhase(\seqx)$, $2^{\numberOfIterations+1}$ output of the Golay layer is converted to Cartesian coordinates. The output of \ac{P2C} layer is perturbed by a noise layer which adds noise with variance $\noiseVariance/2$ real and imaginary parts. The transmitted bits are then detected by a \ac{DNN} with $\numberOfLayersAtRX$ layers.

To achieve a reliable communications under the fading channels, it is also possible to extend the \ac{OFDM-AE} shown in  \figurename~\ref{fig:txrxae}\subref{fig:autoencoder} with several other layers or train it with the existing methods. One approach is to perform the training by including a distortion layer that models the behavior of fading. For example, in \cite{Felix_2018}, it was observed that the \ac{AE} shifts the center of the constellations to superimpose the pilot information on data symbols.  In \cite{Zhao_2018}, a different case for an \ac{OFDM} transmitter is investigated and noted that the receiver complexity can be high and the training can take a longer duration. To overcome these issues, another network between the subcarrier de-mapping and the \ac{DNN} at the receiver along with a two-stage training is proposed. Another approach is to transmit a fixed \ac{OFDM} symbol along with the \ac{OFDM-AE} symbol and let the \ac{DNN} at the receiver perform joint channel estimation and symbol detection \cite{Ye_2018}. 
Given the availability of possible extensions, for the scope of this work, we focus our investigation on the Golay layer for an \ac{AWGN} channel during the training.

\section{Numerical Results}
\label{sec:numerical}
\begin{table}[]
	\centering
	\caption{Layout of the autoencoder}

	\begin{tabular}{cll}
		\hline
		\multicolumn{1}{l}{}                    & Layer               & Output Node \\ \hline
		\multirow{7}{*}{TX}                     & Input                     & 9 (binary)  \\
		& Dense (Batchnorm+ReLU)    & 100         \\
		& Dense (Batchnorm+ReLU)    & 100         \\
		& Dense (Batchnorm+ReLU)    & 100         \\
		& Dense (Clipping layer)    & 11          \\
		& Golay layer               & 64          \\
		& Polar-to-Cartesian           & 64          \\ \hline
		\multicolumn{1}{l}{CH}                  & Noise layer               & 64          \\ \hline
		\multicolumn{1}{l}{\multirow{4}{*}{RX}} & Dense (Batchnorm+ReLU)    & 1000        \\
		\multicolumn{1}{l}{}                    & Dense (Batchnorm+ReLU)    & 1000        \\
		\multicolumn{1}{l}{}                    & Dense (Batchnorm+softmax) & 512         \\
		\multicolumn{1}{l}{}                    & Classification            & 1 (integer) \\ \hline
	\end{tabular}

\label{table:layoutAE}
\end{table}

Assume that $\numberOfInfoBits=9$  bits need to be transmitted. In this case, the transmitter needs to generate $2^9$ distinct \acp{CS} based on the information bits and the receiver should be able distinguish them to receive the information bits. Let $\numberOfIterations=5$, $\deviationPhase=2\pi$, $\seqPermutationCompShift=(1,2,3,4,5)$.  Since the length of each \ac{CS} is $2^5=32$ under these settings, $32$ subcarriers are used for the transmission. Therefore, the \ac{SE} is $9/32$ bit/second/Hz. 

We consider two \ac{OFDM-AE} designs with  $\deviationAmplitude=1$ and  $\deviationAmplitude=0$. The layer information at the transmitter and receiver are provided in \tablename~\ref{table:layoutAE} for both \acp{OFDM-AE}. At the transmitter, the information bits are first processed by three dense layers with batchnorm and \acp{ReLU} and one dense layer with two parallel clipping layers as in \figurename~\ref{fig:txrxae}\subref{fig:autoencoder}. We set the parameters of the clipping layer  such that  $-2\le\scaleEexpBatch[\indexIteration][\indexSampleInBatch]\le1$ for $\indexIteration=1,2,\mydots,5$ and $|\angleexpAllBatch[\indexIteration][\indexSampleInBatch]|\le1$ for $\indexIteration=0,1,\mydots,5$. At the Golay layer,   $\scaleEexpBatch[0][0]$ is calculated based on the condition given in \eqref{eq:normalizationCoef} for the normalization, and $2^5$ outputs for the amplitude function and $2^5$ outputs for the phase function are calculated for the $\indexSampleInBatch$th sample in the minibatch. The \ac{P2C} layer converts this information to Cartesian coordinates. During the training, we set the noise variance on the real and imaginary parts to be $\noiseVariance/2=1/2$. At the receiver, the corresponding subcarriers are processed with two dense layers with batchnorm and \acp{ReLU} and one dense layer. The last dense layer of the \ac{AE} consists of softmax as activation function and the last layer is a classification layer which  returns the index of the output node with the maximum value. For example, if the information bits are (0,0,0,0,0,0,0,0,0), the first element of softmax layer should be closer to 1 while other 511 elements are near 0. Then, the classification layer returns 1. During the training, our batch size is $\sizeofBatch=5120$ and the $\lossFunction$ is the binary cross-entropy function. The learning rate is set to 0.0001 We train the \ac{AE} by using MATLAB Deep Neural Network Toolbox and utilize NVIDIA GeForce GTX 1060. For the sake of comparison, we consider polar code with  length 32 with \ac{BPSK} modulation. We set the design \ac{SNR} as 3 dB and consider \ac{SIC} at the receiver.

\begin{figure}[]
	\centering
	{\includegraphics[width =3.4in]{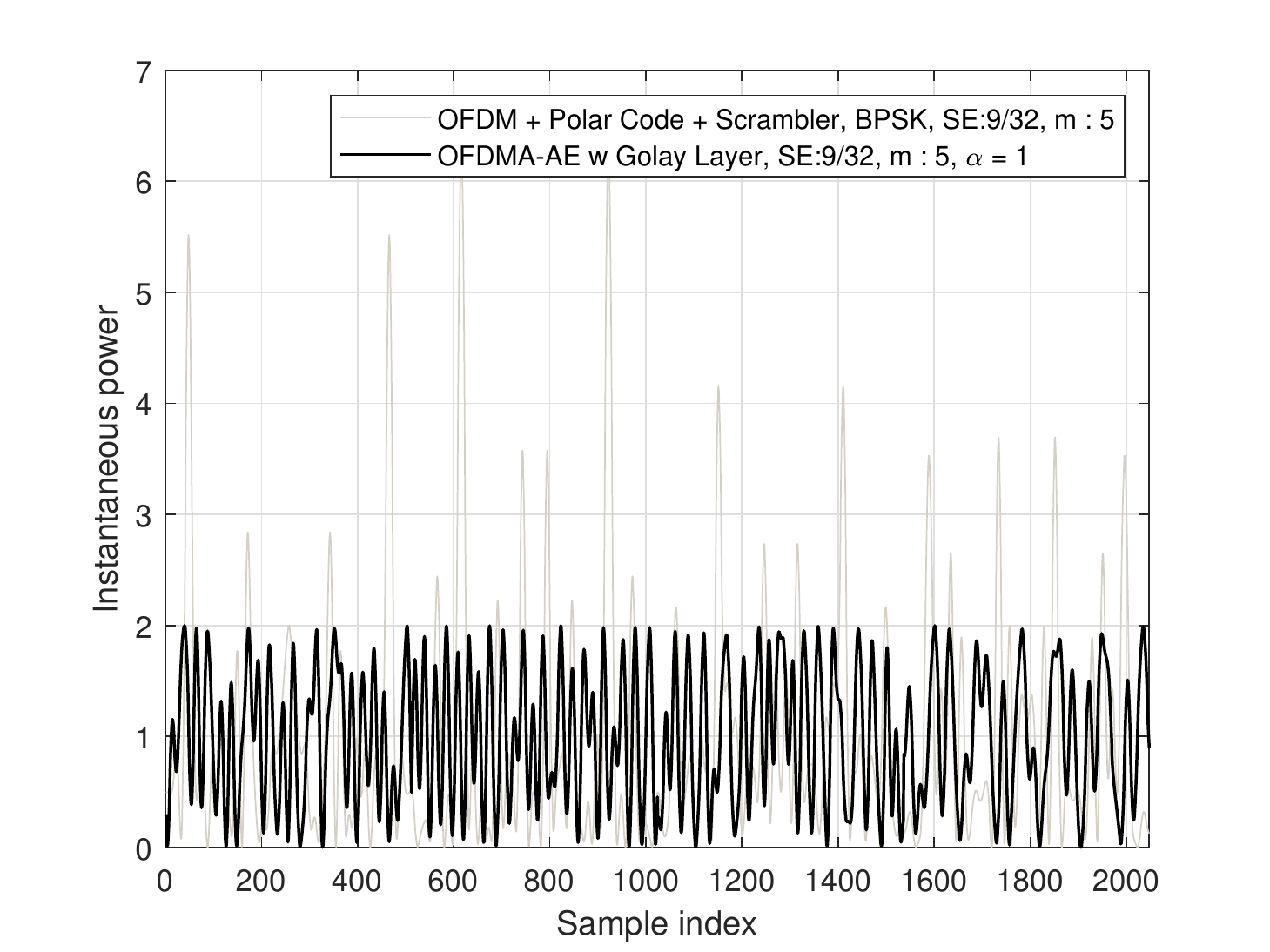}
	}
	\caption{Temporal characteristics.}
	\label{fig:temp}
\end{figure}
\begin{figure}[]
	\centering
	{\includegraphics[width =3.4in]{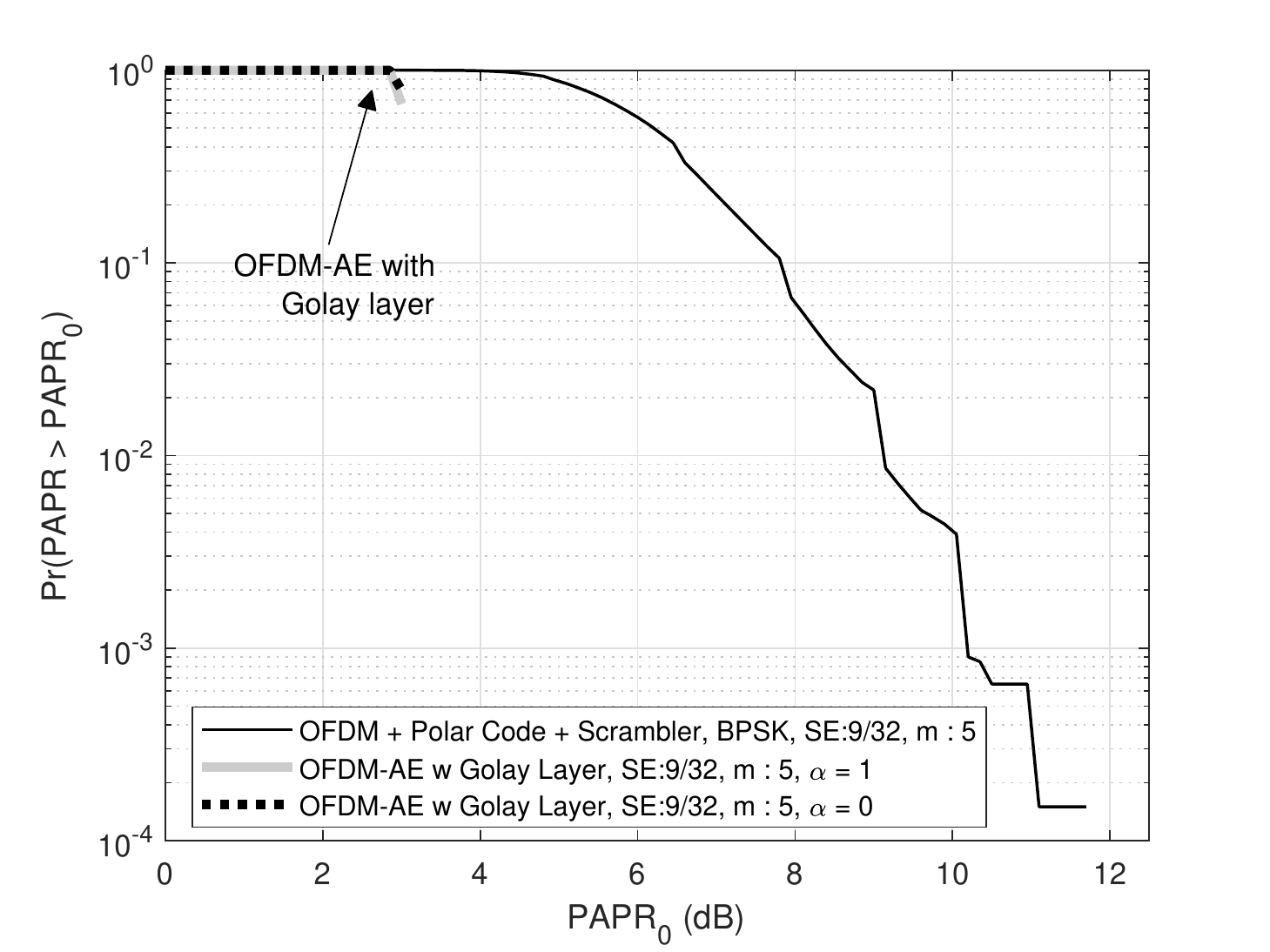}
	}
	\caption{\ac{PAPR} distributions.}
	\label{fig:papr}
\end{figure}
\subsection{Peak-to-Average Power Ratio}
In \figurename~\ref{fig:temp}, we provide the temporal characteristics of five randomly generated \ac{OFDM} symbols (without \ac{CP}) based on Polar code (with a scrambler to reduce \ac{PAPR}) and compare it with the \ac{OFDM-AE} symbols designed for $\deviationAmplitude=1$. As clearly seen, the instantaneous power for the \ac{OFDM-AE} symbols never exceeds 2 (i.e., approximately 3 dB) while the \ac{OFDM} symbols with traditional encoding can be peaky which may require a large power back-off at the transmitter. The \ac{PAPR} distributions are compared in \figurename~\ref{fig:papr}. The \ac{PAPR} gain at the 90th percentile is approximately 6 dB as the Golay layers at \acp{OFDM-AE} limit the \ac{PAPR} to be less than or equal to $3$~dB.

\begin{figure*}[t]
	\centering
	\subfloat[$\deviationAmplitude=1$.]{\includegraphics[width =3in]{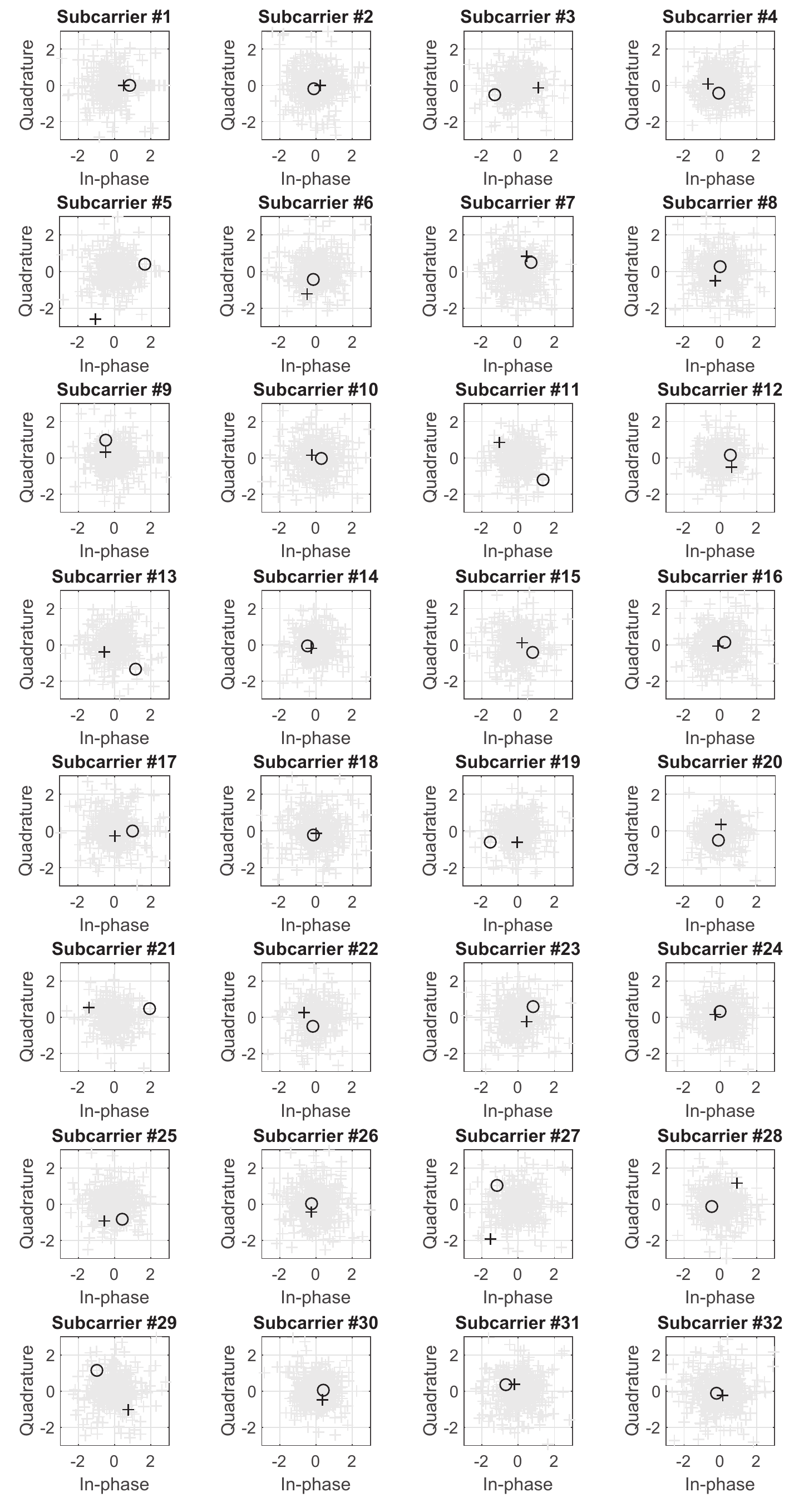}
		\label{subfig:alpha1}}~~~~~~
	\subfloat[$\deviationAmplitude=0$.]{\includegraphics[width =3in]{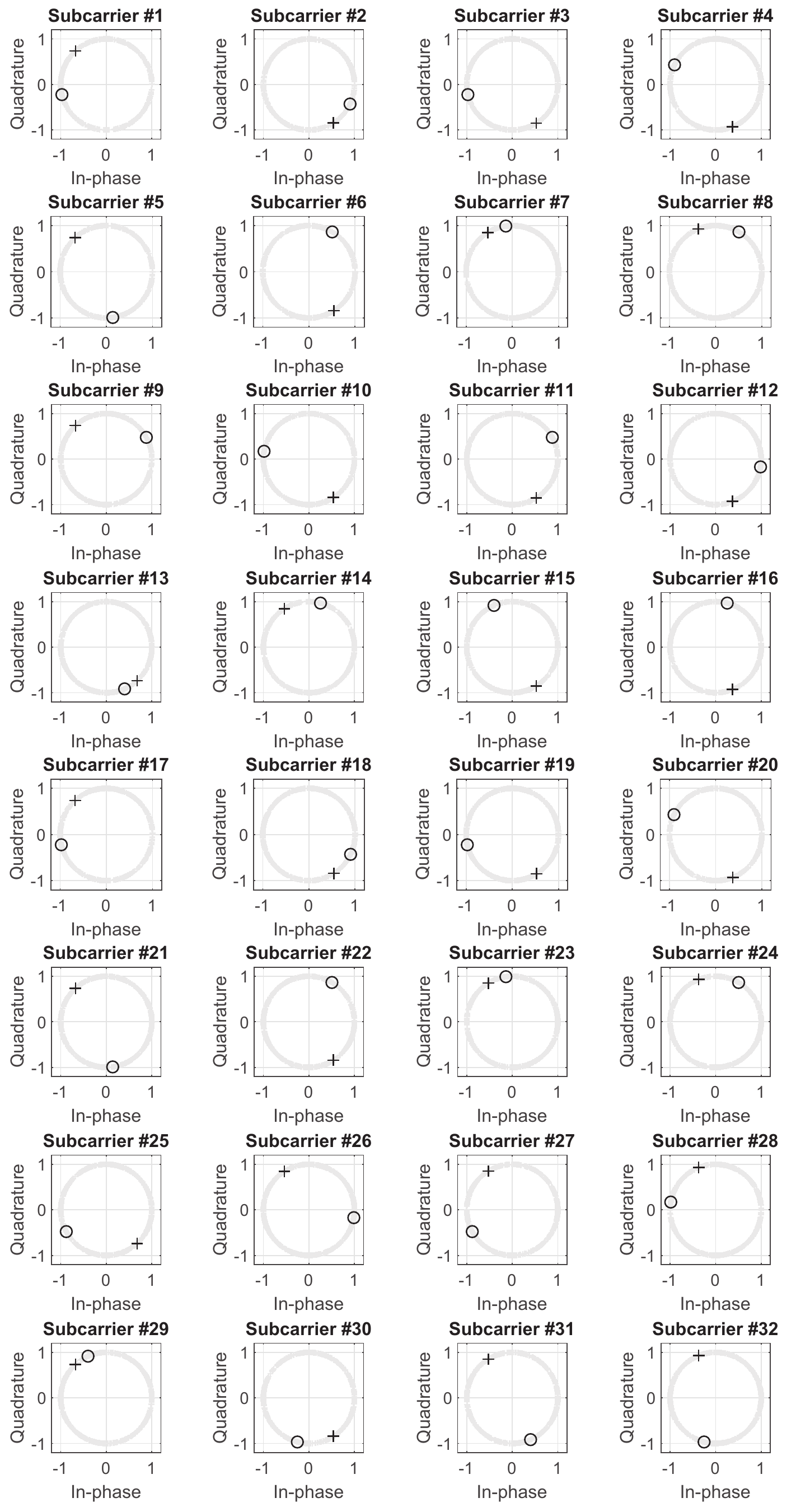}
		\label{subfig:alpha0}}
	\caption{Constellations on the subcarriers (+: \ac{CS} for information bit sequence (0,0,0,0,0,0,0,0,0), o: \ac{CS} for information bit sequence (0,0,0,0,0,0,0,0,1)).}
	\label{fig:const}
	\vspace{-2mm}
\end{figure*}
\subsection{Constellation}
\begin{figure}[]
	\centering
	{\includegraphics[width =3.4in]{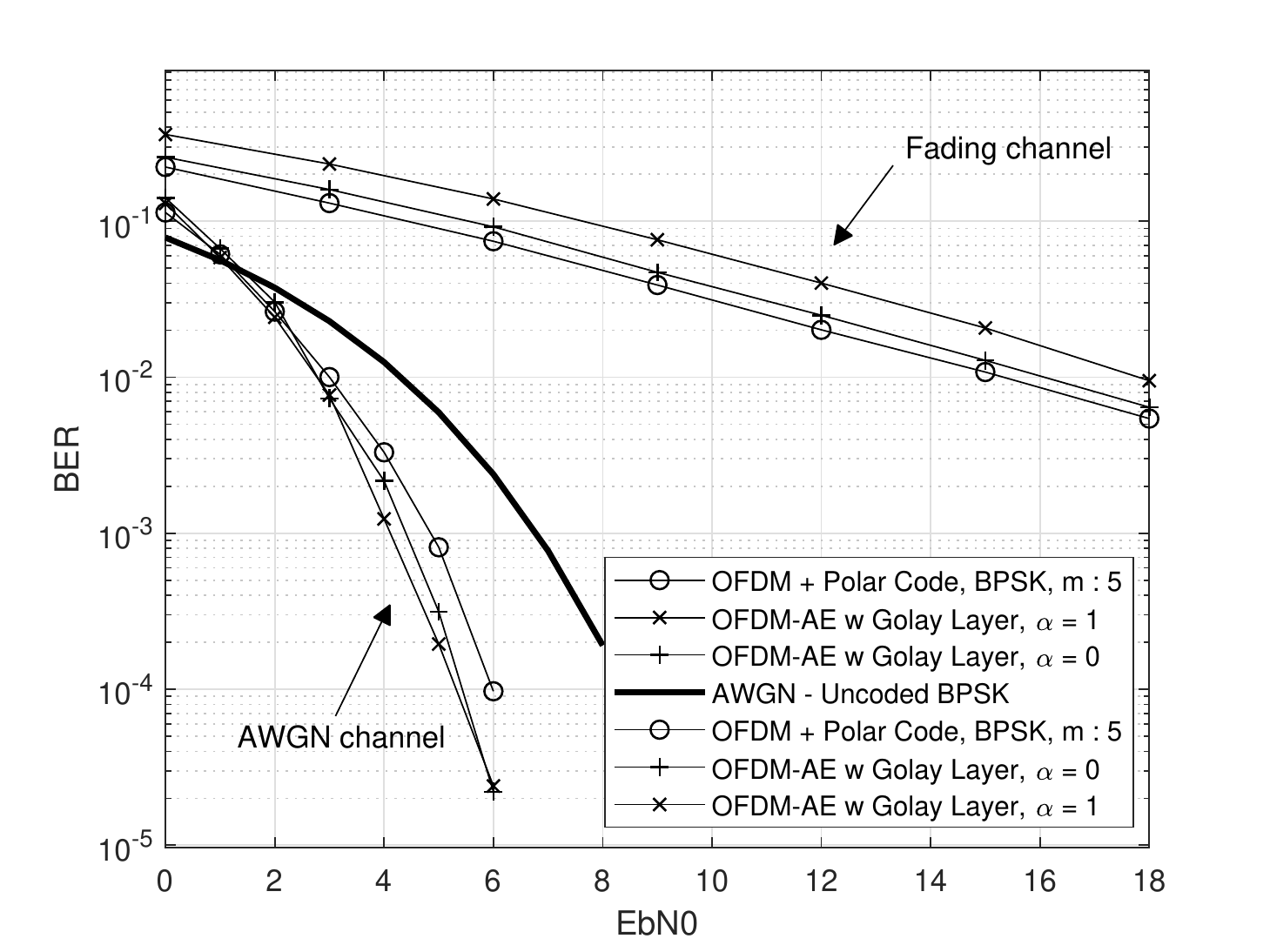}
	}
	\caption{\ac{BER} results.}
	\label{fig:ber}
	\vspace{-3mm}
\end{figure}

In \figurename~\ref{fig:const}, we provide the constellation on each subcarrier obtained for \ac{OFDM-AE} for $\deviationAmplitude=1$ and $\deviationAmplitude=0$. As opposed to the handcrafted designs for \acp{CS} (e.g., \cite{budisin_2018,Sahin_2018}), the \ac{OFDM-AE} does not follow any of the traditional constellations  such as $M$-\ac{QAM} for \acp{CS}.
This result is expected as the \ac{OFDM-AE} does not make isolated symbol-level decisions. 
 In \figurename~\ref{fig:const}, we mark the elements of learned \acp{CS} for two different information bit sequences with + and o. For $\deviationAmplitude=0$, we observe that some of the  values on the subcarriers are very close to each other (e.g., subcarrier \#1, \#2, \#14, \#18) while or there exists points on completely on opposite quadrants (e.g., subcarrier \#3, \#11, \#21, \#29). Nevertheless, it is hard to state that the trained \ac{OFDM-AE} exploits this property to achieve relatively good error rate performance shown in \figurename~\ref{fig:ber} for \ac{AWGN} channel. For $\deviationAmplitude=0$, the constellation on each subcarrier is constrained to be on the unit circle. However, the positions of the elements of the learned \ac{CS} does not appear to follow the same pattern for the case with $\deviationAmplitude=1$.

\vspace{-1mm}
\subsection{Error Rate}
One natural question is if \ac{OFDM-AE} can still perform well  with a Golay layer. Although it is not trivial to provide a definite answer to this question, our numerical results in \figurename~\ref{fig:ber}  demonstrate that the \ac{OFDM-AE} with a Golay layer and \ac{OFDM} with Polar coding can perform similar (within the range of 1.5 dB) under at least aforementioned simulation settings. 
For this comparison, we consider the same spectral efficiency for all schemes.
As shown in \figurename~\ref{fig:ber}, the \ac{SNR} gains for \acp{OFDM-AE} with $\deviationAmplitude=1$ and $\deviationAmplitude=0$  are approximately 1 dB and 0.75 dB, respectively, at 1e-3 \ac{BER} as compared to the \ac{OFDM} with Polar coding under \ac{AWGN} channel.
The origin of the gain is the joint design of modulation and coding through \eqref{eq:realPartReduced}  and \eqref{eq:imagPartReduced}. For example,  \eqref{eq:imagPartReduced} reduces to the first-order RM code within the second-order RM code for \ac{PSK} constellation \cite{davis_1999}.
The \ac{OFDM-AE} with $\deviationAmplitude=1$  performs slightly better than the one with $\deviationAmplitude=0$ since  $\deviationAmplitude=1$  allows the backpropagation to exploit the amplitude of the elements of the \acp{CS}. For the fading channel, we assume a flat fading channel and consider single-tap frequency domain \ac{MMSE} equalization  with ideal channel frequency response. In this case, the performance of the \ac{OFDM} with polar code is  better than that of \acp{OFDM-AE}. This result is expected as the \ac{SIC} exploits the soft bits and the \acp{OFDM-AE} are trained under the \ac{AWGN} channel. Nevertheless, the \ac{PAPR} gain of \acp{OFDM-AE} with the Golay layer is still much larger than the \ac{SNR} gains in this scenario. Another observation from \figurename~\ref{fig:ber} is that the \ac{OFDM-AE} with $\deviationAmplitude=0$ is superior to the one with $\deviationAmplitude=1$ under the fading channel. This indicates that the identical energy for the elements of the learned \ac{CS} can help to achieve better \ac{BER} under the fading channel.



\vspace{-1mm}
\section{Concluding Remarks}\label{sec:conclusion}
In this study, we propose a method that limits the \ac{PAPR} of an \ac{OFDM-AE} symbol to be less than or equal to 3 dB
 without introducing a regularization term to the cost function used in the training. 
It relies on the modification of the differentiable functions that lead to a complex \ac{CS} based on Theorem~\ref{th:reduced}. 
Based on our preliminary results, an \ac{OFDM-AE} with a Golay layer can provide an acceptable \ac{BER} performance. Under our simulation assumptions, we obtained 1 dB SNR gain for error rate at 1e-3 for \ac{AWGN} channel and 6 dB PAPR gain at the 90th percentile as compared to OFDM with a polar code. 

The proposed concept needs to be studied further to achieve a more flexible scheme.
For example, we limit our investigation in this study to only a small number of information bits (i.e., the \ac{SE} is only 9/32 bit/second/Hz) because of the high-complexity of \ac{AE}. Hence, the methods that can give a larger \ac{SE} still need to be investigated under transmitter/receiver complexity  and power consumption constraints. In this direction, the layers at the \acp{DNN} may need to be revised. 
A multiple parallel Golay layer can also be useful for adjusting the \ac{PAPR} level while allowing a larger number of information bits.
Another extension is to develop advanced Golay layers. For example, Theorem~\ref{th:reduced} allows the set of $\{\separationGolay[\indexIteration]|\numberOfIterations=1,2,\mydots,\numberOfIterations\}$ to be arbitrary parameters, which can extend the definition of the basic Golay layer.
Evaluating the scheme through a practical setup is also another angle that can be explored.

\bibliographystyle{IEEEtran}
\bibliography{gmAI}

\end{document}